\documentclass[runningheads,a4paper]{llncs}
\usepackage{cite}
\usepackage{amsmath,amssymb,amsfonts}
\usepackage{algorithmic}
\usepackage{graphicx}
\usepackage{textcomp}
\usepackage{bm}
\usepackage{subfigure}
\usepackage{caption,setspace}

\begin{document}

\mainmatter
\title{Region-Manipulated Fusion Networks for Pancreatitis Recognition}

\author{Jian Wang\inst{1} \and Xiaoyao Li\inst{2}  \and Xiangbo Shu\inst{1,\thanks{Corresponding Author: shuxb@njust.edu.cn}} \and Weiqin Li\inst{2} }

\institute{School of Computer Science, Nanjing University of Science and Technology, No.200, Xiaolinwei Road, Nanjing 210094, PR China. \and
Surgical Intensive Care Unit (SICU), Department of General Surgery, Jinling Hospital, Medical School of Nanjing University, Nanjing 210094, PR China.\\
}

\toctitle{Region-Manipulated Fusion Networks for Pancreatitis Recognition}
\tocauthor{Jian Wang et al.}
\maketitle

\begin{abstract}
  This work first attempts to automatically recognize pancreatitis on CT scan images. However, different form the traditional object recognition, such pancreatitis recognition is challenging due to the fine-grained and non-rigid appearance variability of the local diseased regions. To this end, we propose a customized Region-Manipulated Fusion Networks (RMFN) to capture the key characteristics of local lesion for pancreatitis recognition. Specifically, to effectively highlight the imperceptible lesion regions, a novel region-manipulated scheme in RMFN is proposed to force the lesion regions while weaken the non-lesion regions by ceaselessly aggregating the multi-scale local information onto feature maps. The proposed scheme can be flexibly equipped into the existing neural networks, such as AlexNet and VGG. To evaluate the performance of the propose method, a real CT image database about pancreatitis is collected from hospitals \footnote{The database is available later}. And experimental results on such database well demonstrate the effectiveness of the proposed method for pancreatitis recognition.
\end{abstract}


\section{Introduction}
\label{sec:introduction}

  The physiological and pathological changes of the pancreas are closely related to life. Acute Pancreatitis (AP) is one of the most common gastrointestinal disorder requiring inpatient admission all over the world~\cite{yadav2013epidemiology}. A large part of the patients (15$\sim$20\%) develop severe AP, which is characterized by persistent organ failure, resulting in significant morbidity and mortality~\cite{banks2013classification,cavestro2015single,nawaz2013revised}. Severe AP has a higher mortality rate, which has attracted widespread attention in the last ten years. Therefore, it is urgent to diagnose this dangerous diseases accurately in the critical early phase of the AP.

  As we know, Computed Tomography (CT) is a widely-used technology that provides a common way for the diagnose of AP in the early phase. Generally, a radiologist makes a primary diagnosis, the clinician can then make a definite diagnosis via the radiologist's decision, laboratory data and clinical manifestation. However, there are some shortcomings in such diagnose approaches. Firstly, radiologists are difficult to keep absolute objectivity due to limitations of their own knowledge and experience. Secondly, reading CT is a time-consuming and labor consuming work. Finally, the size of early lesions is usually small and the characteristics are generally not obvious enough, which easily leads to missed diagnosis.
%\begin{figure*}[t]
%\begin{center}
%%\fbox{\rule{0pt}{2in} \rule{0.9\linewidth}{0.png}}
%  \includegraphics[width=1\linewidth]{1.pdf}
%\end{center}
%\vspace{2mm}
%   \caption{The architecture of the proposed networks. The architecture has three networks consists of a main-network and two sub-networks. The input image consists of three scales. The feature map $FS_1$ of sub-network 1 and feature map $FM_1$ from FCN-M1 of main-network are fused to $FF_1$ in fusion 1 strategy. The feature map $FS_2$ of sub-network 2 and feature map $FM_2$ from FCN-M2 of main-network are fused to $FF_2$ in fusion 2 strategy. FCN-M3 of main-network takes $FF_2$ as input and outputs $FM_3$. Finally, the FC of main-network takes $FM_3$ as input and outputs predicted results.}%{}
%\label{fig:main}
%%\label{fig:onecol1}
%\end{figure*}

\begin{figure}[t]
\begin{center}
%\fbox{\rule{0pt}{2in} \rule{0.9\linewidth}{figure1.jpg}}
  \includegraphics[width=1\linewidth]{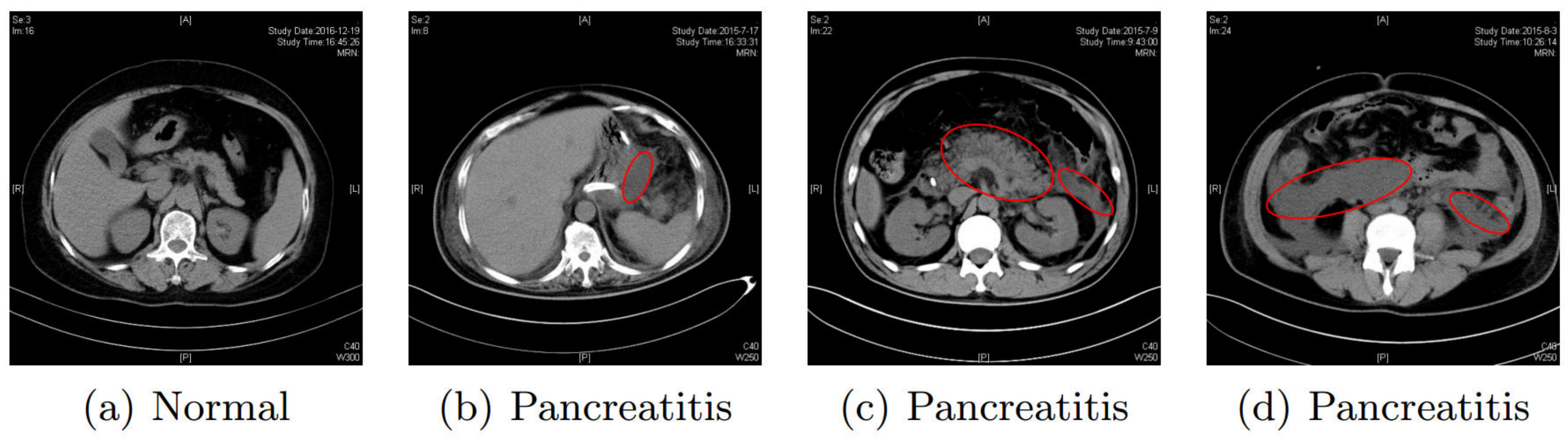}
\end{center}
  \vspace{2mm}
 \caption{Comparisons of different abdominal CT scan images. The lesion regions are outlined in red circles. It is obvious that the lesion area has only a small local area in the whole image and most area of the image is normal. (a) normal abdominal CT scan image. (b) fluid collection in omental bursa and posterior gastric space. (c) pancreatic swelling and fluid collection in the left anterior pararenal space. (d) fluid collection in paracolic space.} %
 \label{fig:first}
\end{figure}

  Recently, inspired by the tremendous development of computer vision and machine learning, intelligent medical diagnosis~\cite{girshick2015fast} explores to automatically recognize whether there are certain diseases on CT scan images. This can not only greatly improve the efficiency of diagnosis, but also avoid the human subjectivity and even misdiagnosis as much as possible. At present, deep learning has outperformed traditional machine learning algorithms in various vision problems, e.g., natural image classification~\cite{krizhevsky2012imagenet}, object detection~\cite{girshick2015fast}, and face recognition~\cite{sun2015deeply}, etc. The excellent performance of deep learning in computer vision has attracted the attention of medical image processing researchers. However, most existing deep learning models are designed for natural images, which consider the whole image instead of the local area. In contrast, medical images have large intra-class variation and inter-class ambiguity. In addition, in the medical image of a patient, lesion area providing valuable information is usually small, while most areas are normal. Undoubtedly, faced with this challenging task, several early-bird researchers have designed various architectures of deep networks for some specific tasks, such as diabetic retinopathy detection~\cite{gulshan2016development}, skin cancer classification~\cite{esteva2017dermatologist}, and lung nodule classification~\cite{golan2016lung}, etc.

  Inspired by the above work, we first attempts to automatically recognize pancreatitis on abdominal CT scan images by leveraging the superior power of deep learning. However, according to the investigation with the collaboration of doctors in the early stage, image recognition of pancreatitis has to face with several specific challenges compared with existing deep learning based medical analysis methods. As shown in  Figure 1, three great challenges are summarized as follows, (1) {\bf fine-grained recognition}. The lesion area only providing the valuable information is very small. (2) {\bf intra-class variation and inter-class ambiguity}. The intra-class variation between images from the same class is common, and the inter-class difference between images from different classes is ambiguous. (3) {\bf Non-rigid object}. The lesion region doesn't have a certain shape and size.

  After above analysis, we propose a customized Region-Manipulated Fusion Networks (RMFN) to automatically recognize pancreatitis on the abdominal CT scan images in this work. First, to solve this fine-grained recognition, we adopt a deep network architecture to learn the high-level representations of the CT images. Second, to avoid intra-class variation and inter-class ambiguity, we also capture the local discriminative information from the local view by taking the small regions as the input of the subnetworks. Finally, to highlight the non-rigid lesion, a novel region-manipulated scheme is proposed to force the lesion regions while weaken the non-lesion regions by ceaselessly aggregating the multi-scale local information onto feature maps. Specifically, we fuse multi-scale feature maps from the FCN of main-network and sub-networks, and then the fused feature map is input to the next FCN of main-network. To well evaluate the performance of the proposed method, we collect a real-world CT image database about pancreatitis from hospitals, and conduct experiments on such database to well demonstrate the effectiveness of the proposed method.

  In summary, the main contributions of this work are two-fold: (1) To  the  best  of  our  knowledge, our work is the first time to leverage a novel CNN architecture to solve the problem of pancreatitis recognition, and contribute a database of abdominal CT scan images on the community of medical image analysis. (2) We propose a novel region-manipulated scheme to enhance small yet discriminative information regions of images. The proposed scheme can be flexibly equipped into the existing neural networks to improve the performance of image classification and object recognition in the community of computer vision.

\begin{figure}[t]
\begin{center}
%\fbox{\rule{0pt}{2in} \rule{0.9\linewidth}{0.png}}
  \includegraphics[width=1\linewidth]{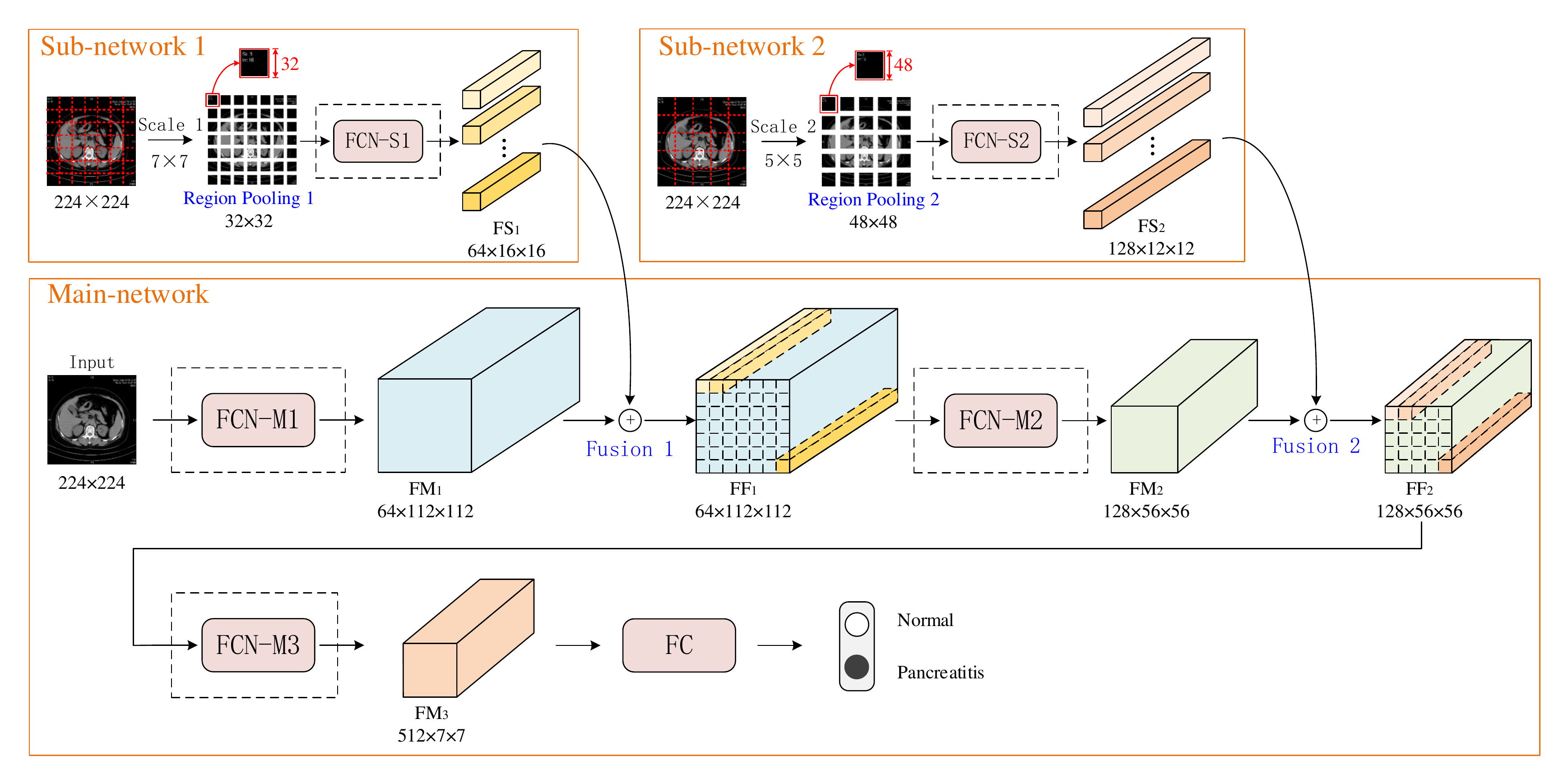}
\end{center}
\vspace{2mm}
   \caption{The architecture of the proposed networks. The architecture has three networks consists of a main-network and two sub-networks. The input image consists of three scales. The feature map $FS_1$ of sub-network 1 and feature map $FM_1$ from FCN-M1 of main-network are fused to $FF_1$ in fusion 1 strategy. The feature map $FS_2$ of sub-network 2 and feature map $FM_2$ from FCN-M2 of main-network are fused to $FF_2$ in fusion 2 strategy. FCN-M3 of main-network takes $FF_2$ as input and outputs $FM_3$. Finally, the FC of main-network takes $FM_3$ as input and outputs predicted results.}%{}
\label{fig:main}
%\label{fig:onecol1}
\end{figure}

\section{Related works}

  In this section, we give a brief review to deep learning and medical image analysis with deep learning.

\subsection{Deep learning in computer vision}
  As a feature learning method, deep learning~\cite{lecun2015deep} can transform raw data into a higher level and more abstract representation. These abstract representations are advantageous to learn essential features of images~\cite{zhou2016learning}. Therefore, deep learning has brought a breakthrough in many fields of computer vision, such as face recognition~\cite{parkhi2015deep,liu2017sphereface}, image classification~\cite{akata2014good,wang2017residual,szegedy2015going}, object detection~\cite{girshick2015fast,ren2015faster},  image segmentation~\cite{chen2017deeplab,badrinarayanan2017segnet,chen2018encoder}, etc.

  Many variants of deep learning models have been proposed for the last five years, such as AlexNet~\cite{krizhevsky2012imagenet}, VGG~\cite{simonyan2014very}, GoogleNet~\cite{szegedy2015going}, ResNet~\cite{he2016deep}, and so on. In the image classification field, Krizhevsky et al. proposed AlexNet and achieved great improvement compared with the other methods on the ILSVRC-2012 classification task~\cite{krizhevsky2012imagenet}. He et al. proposed a Deep Residual Learning model and achieved encouraging results on the ILSVRC-2015 classification task~\cite{he2016deep}. In the object detection field, Girshick proposed a Fast R-CNN to speed up the detection and improve detection accuracy on the image detection filed~\cite{girshick2015fast}. In the image segmentation field, Chen et al. presented a Deeplab for semantic image segmentation and achieved the-state-of-art results at the PASCAL VOC-2012 semantic image segmentation task~\cite{chen2017deeplab}.

  The excellent performance of CNN in computer vision attracted great attention of medical image researchers. More and more medical image researchers are applying CNN to solve the tasks of medical images, such as classification~\cite{esteva2017dermatologist,ren2018adversarial,zhang2018skin}, detection~\cite{liao2019evaluate,zhu2018deeplung}, segmentation ~\cite{farag2017bottom,man2019deep,yu2018recurrent}, etc.

\subsection{Medical image analysis with deep learning}
  There are a large number of medical image data in medical institutions, such as CT scan images. These medical images usually contain a great deal of potential information~\cite{lambin2012radiomics}. Therefore, we need to make full use of these large amounts of data and potential information. The diagnosis usually relies on doctors' expert knowledge that may be time consuming and information loss. Recently, the rapid development of deep learning provides a feasible solution for the development of medical automation.

  Some medical image researchers began to study many specific diseases. For example, Esteva et al. used deep learning methods to classify skin cancer diseases, and the diagnostic accuracy was comparable to that of human doctors~\cite{esteva2017dermatologist}. Gulshan et al. employed the deep learning methods to detect the diabetic retinopathy ~\cite{gulshan2016development}. Gao et al. applied deep learning methods to classify interstitial lung diseases imaging patterns on CT images~\cite{gao2016holistic}. However, the performance of deep learning lies on the larger scale database. In fact, there is only a few of high-quality annotated database of medical images available. Usually, once a large-scale medical image database is reported, there will be emerging several works to solve various problems on this database. For example, lung nodules on CT scans is collected by Armato et al.~\cite{armato2011lung}, subsequently, there emerged a lot of studies on recognizing lung nodules with deep learning methods~\cite{golan2016lung,setio2016pulmonary,shin2016deep}.

  In the field of medical analysis, many researchers focus on the network structures to make them more suitable for specific diseases. Gao et al. proposed a novel framework that combines CNN and RNN for cataract grading and achieved the state-of-the-art performance~\cite{gao2015automatic}. Kamnitsas et al. proposed a dual pathway architecture to process the input images at multiple scales~\cite{kamnitsas2017efficient}. They make full use of multi-scale contextual information and thus achieve a satisfactory performance. Inspired by above works, we also attempt to re-design the deep learning architecture to  recognize pancreatitis on abdominal CT scan images in this work.

\section{The proposed method}
\label{}
  In order to recognize pancreatitis accurately, we propose a customized Region-Manipulated Fusion Networks (RMFN) by ceaselessly aggregating the multi-scale local information onto feature maps. Here, the proposed region-manipulated scheme can be incorporated into the existing neural networks, such as AlexNet~\cite{krizhevsky2012imagenet} and VGG~\cite{simonyan2014very}. In this section, we take VGG11 as the base networks to describe the proposed RMFN.

\subsection{Network architecture}

  The architecture of the proposed RMFN is shown in Figure 2. It can be seen that RMFN consists of a main-network and two sub-networks. Main-network aims to capture the global information, while sub-networks are used to force obtain local discriminative information. Fusion of main-network and sub-networks can enhance the ability of model to highlight the local lesion of abdominal CT images with pancreatitis. Given an image, a fully convolutional network M1 (FCN-M1) takes it as input and outputs the corresponding feature map $FM_1 (a)$. To obtain multi-scale information, we use two other scales (i.e., $G_1\times G_1$ and $G_2\times G_2$ grids, respectively).

  At scale 1 ($\alpha1$ for short), we divide the input image into $G_1\times G_1$ regions $\{R_{\alpha1}^{m,n}\}_{m,n}^{G_1}$ by a Region Pooling 1 strategy and put them into a Fully Convolutional Network S1 (FCN-S1) to output the corresponding feature maps $FS_1 (R_{\alpha1}^{m,n}$). Here, $R_{\alpha1}^{m,n}$ denotes the $m$-row and $n$-column region, $m,n=1,2,...,G_1$.

  Once the feature maps $FS_1 (R_{\alpha1}^{m,n})$ and $FM_1 (a)$ are obtained, we take a fusion mechanism (i.e., fusion 1) to fuse these two features and generate a new feature maps $FF_1 (a)$. The $G_1\times G_1$ local regions of feature maps $FM_1 (a)$ can be represented as follows:

\begin{equation}
M_{\alpha1}^{m,n}: \left\{
\begin{aligned}
(m-1)L_{\alpha1}/2\leq{i}\leq{mL_{\alpha1}/2}\\  %,    m=1,2,...,7
(n-1)L_{\alpha1}/2\leq{j}\leq{nL_{\alpha1}/2}
\end{aligned}
\right.
\end{equation}
  where, $M_{\alpha1}^{m,n}$ is the $m$-row, $n$-column grid region of $FM_1 (a)$, $L_{\alpha1}$ is the size of a grid region $R_{\alpha1}^{m,n}$.  $i,j\in \mathbb N^{*}$.

  After getting the feature map $FF_1 (a)$, the Fully Convolutional Network M2 (FCN-M2) takes it as input and the feature map $FM_2 (a)$ is obtained.
 Similar to the scale 1, at scale 2 ($\alpha2$ for short), we divide the input image into $G_2\times G_2$ equal regions $R_{\alpha2}^{m,n}$ by the Region Pooling 2 strategy and put them into a Fully Convolutional Network S2 (FCN-S2) to output the corresponding feature maps $FS_2 (R_{\alpha2}^{m,n})$, where $m,n=1,2,...,G_2$.

  Once the feature maps $FS_2 (R_{\alpha2}^{m,n})$ and $FM_2 (a)$ are obtained, we take a fusion mechanism (fusion 2) to fuse the two features and generate new feature maps $FF_2 (a)$. The $G_2\times G_2$ corresponding regions of feature maps $FM_2 (a)$ are represented as follows:

\begin{equation}
M_{\alpha2}^{m,n}: \left\{
\begin{aligned}
(m-1)(L_{\alpha2}-\epsilon)/4\leq{i}\leq{m(L_{\alpha2}-\epsilon)/4+\epsilon/4} \\%,    m=1,2,...,5
(n-1)(L_{\alpha2}-\epsilon)/4\leq{j}\leq{n(L_{\alpha2}-\epsilon)/4+\epsilon/4}
\end{aligned}
\right.
\end{equation}
  where, $M_{\alpha2}^{m,n}$ is the $m$ row, $n$ column grid region of $FM_2 (a)$. $L_{\alpha2}$ is the side length of the square grid region $R_{\alpha2}^{m,n}$.  $i,j\in \mathbb N^{*}$.

  After getting feature map $FF_2 (a)$, a Fully Convolutional Network M3 (FCN-M3) takes it as input and the feature map $FM_3 (a)$ is obtained. Finally, the Fully Connected networks (FC) takes feature maps $FM_3 (a)$ as input and outputs two-dimensional vector $(x,y)$. More specially, if $x>y$, it indicates that the person is normal in terms to the pancreas, otherwise pancreatitis patient.

\begin{figure*}
\centering
\subfigure[]{
  \begin{minipage}[b]{0.8\textwidth}
      \includegraphics[width=1\linewidth]{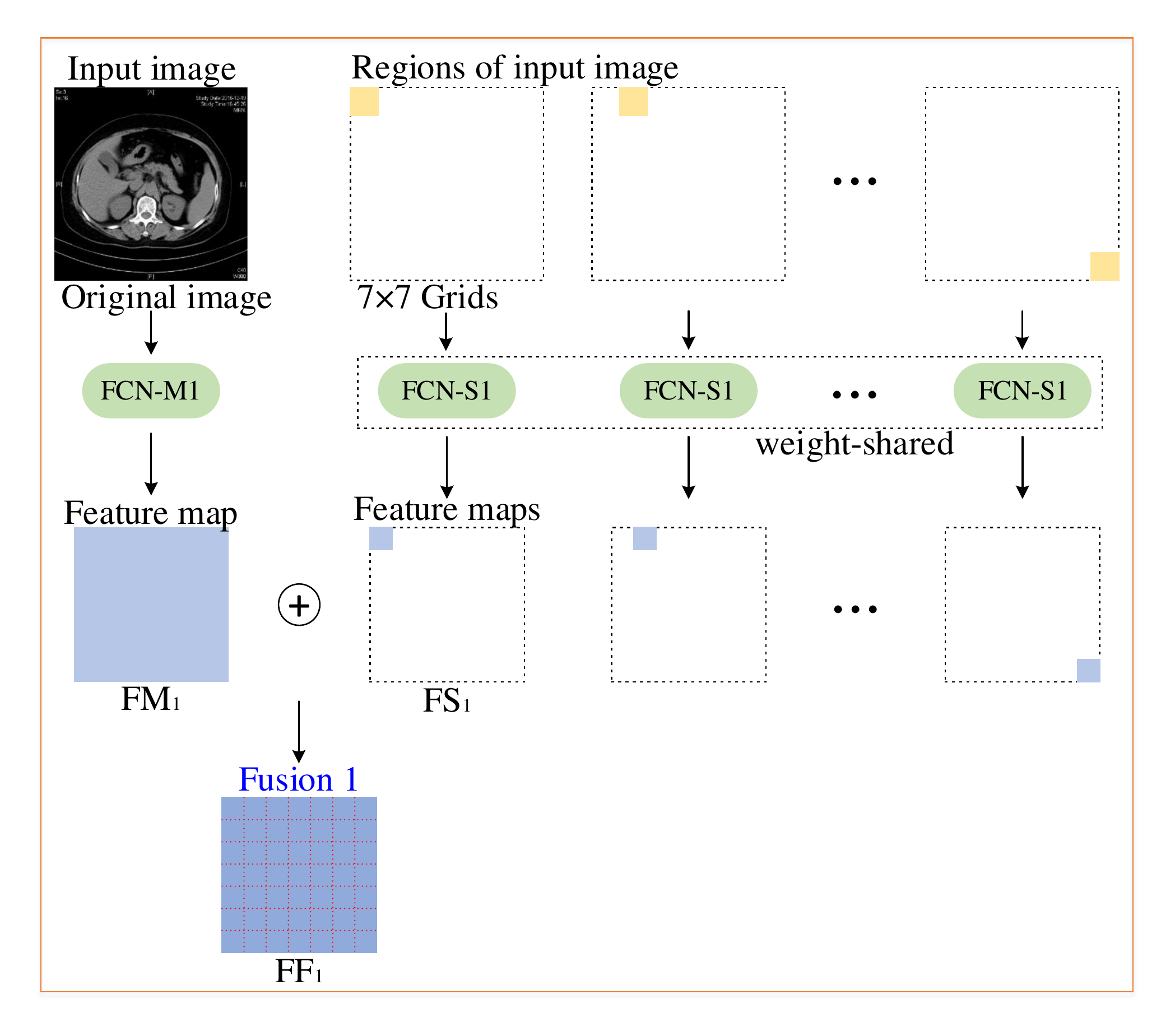}%\\[1mm] %hfill
  \end{minipage}
  }
\subfigure[]{
  \begin{minipage}[b]{0.8\textwidth}
      \includegraphics[width=1\linewidth]{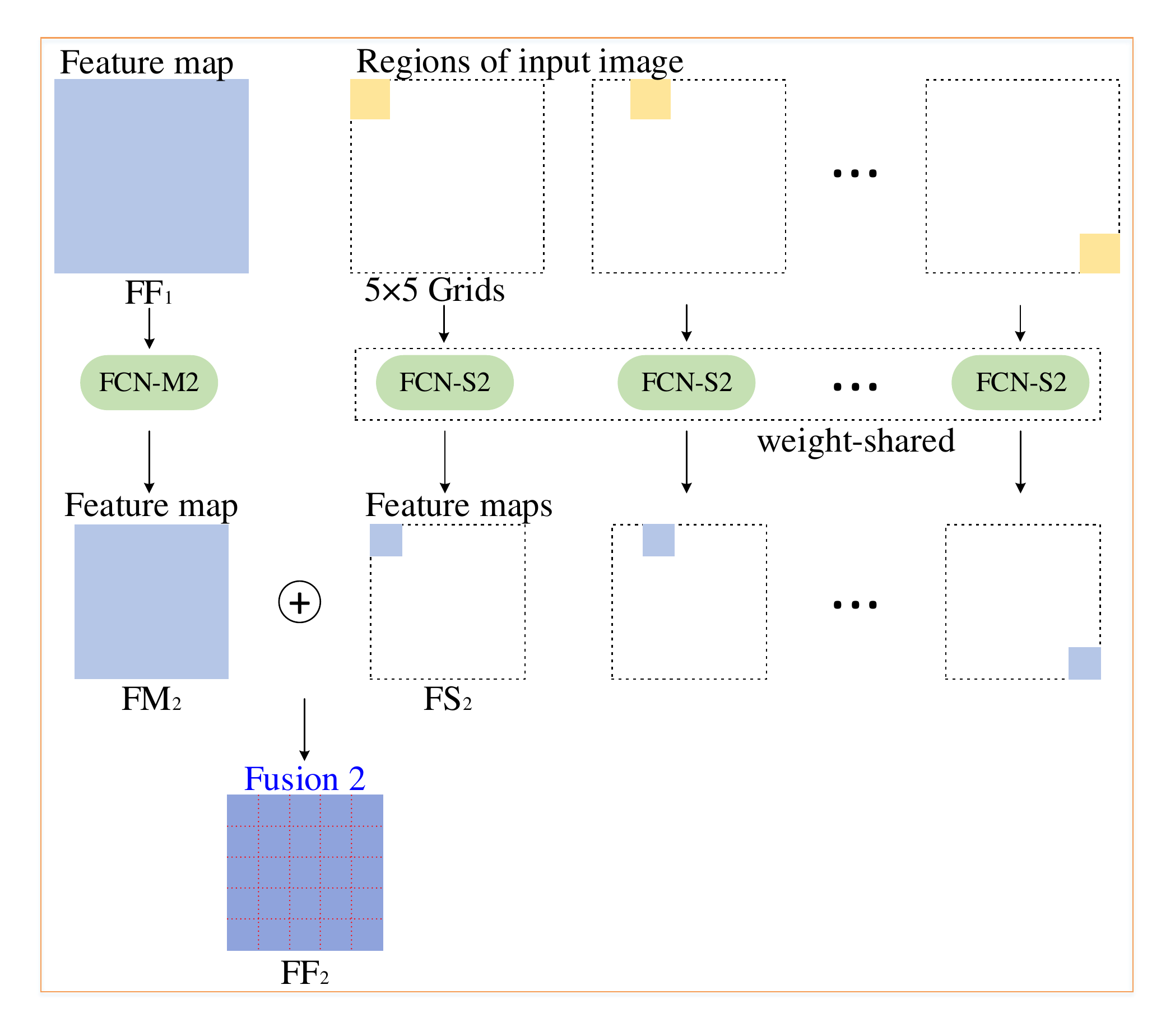}
  \end{minipage}
  }

\vspace{2mm}
	\caption{Two types of Fusion strategies on feature maps. (a) shows that the feature map $FM_1$ of original image obtained by FCN-M1 and the feature map $FS_1$ of Scale 1 (e.g. $7\times7$ grids) image obtained by FCN-S1 are added to feature map $FF_1$. (b) shows that the feature map $FM_2$ of $FF_1$ obtained by FCN-M2 and the feature map $FS_2$ of Scale 2 (e.g. $5\times5$ grids) image obtained by FCN-S2 are added to feature map $FF_2$.}
	\label{fig6}
\end{figure*}
\subsection{Region pooling}
  The texture feature of local region contains important discriminant information. The focus of the paper is to capture texture of non-rigid region. We can strengthen the ability of capture with specific multi-scale grid.

  In order to divide the grid reasonably and obtain multi-scale information, we take a region pooling mechanism to divide the input image into multiple equal regions. The proposed fusion method of this work including two types of strategies: Region Pooling 1 and Region Pooling 2. The strategy of Region Pooling 1 is formulated as:
\begin{equation}
R_{\alpha1}^{m,n}: \left\{
\begin{aligned}
(m-1)L_{\alpha1}\leq{i}\leq{mL_{\alpha1}} \\%,    m=1,2,...,7
(n-1)L_{\alpha1}\leq{j}\leq{nL_{\alpha1}}
\end{aligned}
\right.
\end{equation}
  where $R_{\alpha1}^{m,n}$ ($m,n=1,2,...,G_1$) is the $m$ row, $n$ column grid region of input image, $L_{\alpha1}$ is the size of the square grid region $R_{\alpha1}^{m,n}$, and $i,j\in \mathbb N^{*}$. %(the resolution of every region is $L_{\alpha1}\times L_{\alpha1}$) the length of one side of a square

  Unlike the Region Pooling 1 strategy, the Region Pooling 2 strategy is only used to solve the existence of the overlap area between two adjacent grid regions. Specifically, the method of Region Pooling 2 is formulated as:
\begin{equation}
R_{\alpha2}^{m,n}: \left\{
\begin{aligned}
(m-1)(L_{\alpha2}-\epsilon)\leq{i}\leq{m(L_{\alpha2}-\epsilon)+\epsilon}  \\%,    m=1,2,...,5
(n-1)(L_{\alpha2}-\epsilon)\leq{j}\leq{n(L_{\alpha2}-\epsilon)+\epsilon}   %,    n=1,2,...,5
\end{aligned}
\right.
\end{equation}
%  !!!!!  The two regions overlap each other !!!!!
  where  $i,j\in \mathbb N^{*}$, $R_{\alpha2}^{m,n},m,n=1,2,...,G_2$ is the $m$ row, $n$ column grid region of input image, $L_{\alpha2}$ is size of a square grid region $R_{\alpha2}^{m,n}$, and $\epsilon$ is the width of the overlap area between arbitrary two adjacent grid regions. It is noted that $L_{\alpha2}$ and $\epsilon$ satisfy the following formula:
\begin{equation}
(G_2-1)\epsilon-G_2 L_{\alpha2}+L_0=0
\end{equation}
where $L_0 \times L_0$ is the size of the input image.

\subsection{Fusion strategy}
  In order to make full use of multi-scale feature maps, we take a fusion mechanism to fuse the two features from main-network and sub-network respectively and generate new feature maps. In this work, according to the two types of region pooling strategies, the designed fusion strategy includes two type of versions: Fusion 1 and Fusion 2. First, the Fusion 1 strategy can be formulated as:
\begin{equation}
D_{\alpha1}^{m,n}=\sum\limits_{j=1}^{ L_{\alpha1}/2 }\sum\limits_{i=1}^{ L_{\alpha1}/2 }u_{i,j}^{m,n}+U_{  P_{\alpha1}(i,m),Q_{\alpha1}(j,n) }^{m,n}
\end{equation}
  where $P_{\alpha1}(i,m)=i+(m-1)L_{\alpha1}/2$,  $Q_{\alpha1}(j,n)=j+(n-1)L_{\alpha1}/2$, $u_{i,j}^{m,n}$ is the $i$ row, $j$ column value of the $m$ row, $n$ column region $FS_1 (R_{\alpha1}^{m,n})$. $U_{  P_{\alpha1}(i,m),Q_{\alpha1}(j,n) }^{m,n}$ is the $i$ row, $j$ column value of the region $M_{\alpha1}^{m,n}$. $D_{\alpha1}^{m,n}$ is the whole values of the $m$ row, $n$ column regions $M_{\alpha1}^{m,n}$. Figure 3(a) shows the detailed steps of the Fusion 2 strategy.

  Unlike Fusion 1, Fusion 2 is only used to solve the existence of the overlap area between two adjacent grid regions. Fusion 2 can be formulated as:
\begin{equation}
D_{\alpha2}^{m,n}=\sum\limits_{j=1}^{ L_{\alpha2}/4 }\sum\limits_{i=1}^{ L_{\alpha2}/4 }v_{i,j}^{m,n}+V_{P_{\alpha2}(i,m),Q_{\alpha2}(j,n)}^{m,n}
\end{equation}
  where $P_{\alpha2}(i,m)=i+(m-1)(L_{\alpha2}-\epsilon)/4$,  $Q_{\alpha2}(j,n)=j+(n-1)(L_{\alpha2}-\epsilon)/4$. $v_{i,j}^{m,n}$ is the $i$-row, $j$-column value of the $m$-row, $n$-column region $FS_2 (R_{\alpha2}^{m,n})$. $V_{P_{\alpha2}(i,m),Q_{\alpha2}(j,n)}^{m,n}$ is the $i$-row, $j$ column-value of the corresponding region $M_{\alpha2}^{m,n}$. $D_{\alpha2}^{m,n}$ is the value after fusing the $m$ row and $n$ column regions $M_{\alpha2}^{m,n}$. Figure 3(b) shows the detailed steps of the Fusion 2 strategy.

\subsection{Implementation details}
  In this paper, we take a VGG11 as the base networks to introduce the implementation details of the proposed RMFN. The architecture and configuration of the proposed RMFN is shown in Figure 2 and Table 1, respectively. We can see that the parameter of feature maps $FM_1 (a)$ is $64\times112\times112$, where 64 denotes the number of feature maps, and  $112\times112$ denotes the size of one feature map. Based on the validation for different scales, the parameter $G_1$ of Scale 1 and the parameter $G_2$ of Scale 2 are  set to 7 and 5, respectively. Correspondingly, the parameter $L_{\alpha1}$ of Region Pooling 1 and the parameter $L_{\alpha2}$ of Region Pooling 2 are set to 32 and 48, respectively. In particular, at scale 2, the width $\epsilon$ of the overlap area between two adjacent grid regions is set to 4.

\begin{table*}
		\centering
		\caption{Model configurations of the networks. For example, conv3 and 64 of conv3-64 denote $3\times3$ convolution kernel and the number of feature map, respectively. 4096 of FC-4096 denotes the parameter number of the fully connected layer. }
		\begin{tabular}{|c|c|c|c|c|c|c|c|}
	%        \hline \multicolumn{7}{|c|}{?}\\
	        \hline \multicolumn{2}{|c|}{Original configuration}&\multicolumn{6}{|c|}{Configuration of the proposed network architecture}\\
            %\hline Feature extractor&Classifier&Feature extractor $1\_1$&Feature extractor $1\_2$&Feature extractor 2&Feature extractor 3&Classifier\\
            \hline FCN&FC&FCN-M1&FCN-M2&FCN-M3&FCN-S1&FCN-S2&FC\\
            %extractor& &extractor $1\_1$&extractor $1\_2$&extractor 2&extractor 3& \\
            \hline
	        \hline conv3-64 &FC-4096 &conv3-64 &conv3-128&conv3-256&conv3-64 &conv3-64 &FC-4096\\
	         maxpool  &FC-4096 &maxpool  &maxpool  &conv3-256&maxpool  &maxpool  &FC-4096\\
	         conv3-128&FC-2    &         &         &maxpool  &         &conv3-128&FC-2\\
	         maxpool  &        &         &         &conv3-512&         &maxpool  & \\
	         conv3-256&        &         &         &conv3-512&         &         &\\
	         conv3-256&        &         &         &maxpool  &         &         & \\
	         maxpool  &        &         &         &conv3-512&         &         & \\
	         conv3-512&        &         &         &conv3-512&         &         & \\
	         conv3-512&        &         &         &maxpool  &         &         & \\
	         maxpool  &        &         &         &         &         &         & \\
	         conv3-512&        &         &         &         &         &         & \\
	         conv3-512&        &         &         &         &         &         & \\
	         maxpool  &        &         &         &         &         &         & \\
            \hline
		\end{tabular}
		\label{tab:Margin_settings}
\end{table*}

  In this work, the size of the input  abdominal CT image is set to a $224\times224$. Due to the difference of convolution schemes, the sizes of multi-scale image of different networks are also different. Specifically, when main-network is VGG11 or VGG16, the resolution of multi-scale images is $224\times224$, $48\times48$ ($5\times5$ grids) and $32\times32$ ($7\times7$ grids); when main-network is AlexNet, the resolution of multi-scale images is $224\times224$, $48\times48$ ($5\times5$ grids) and $36\times36$ ($7\times7$ grids). Since the number of VGG16 network parameter is relatively large, which requires much GPU memory, we reduce the number of VGG16 network parameters on a degree, namely the number of parameters is reduced to half.
  We implement the proposed RMFN on PyTorch platform, and choose the stochastic gradient descent (SGD) to update the parameters of the model. The dropout rate, the momentum, batch size and learning rate are set to 0.5 and 0.9, 64 and 0.001, respectively. %}

\section{Experiments}
\label{}
  To evaluate the performance of the proposed method, we collect a real CT image database about pancreatitis from hospitals and conduct experiments on this database. Besides, we also visualize the feature maps of the proposed RMFN and the baseline network models to illustrate the superiority of the proposed region-manipulated scheme.

\begin{figure}[t]%htbp
\begin{center}
%\fbox{\rule{0pt}{2in} \rule{0.9\linewidth}{figure2.jpg}}
  \includegraphics[width=1\linewidth]{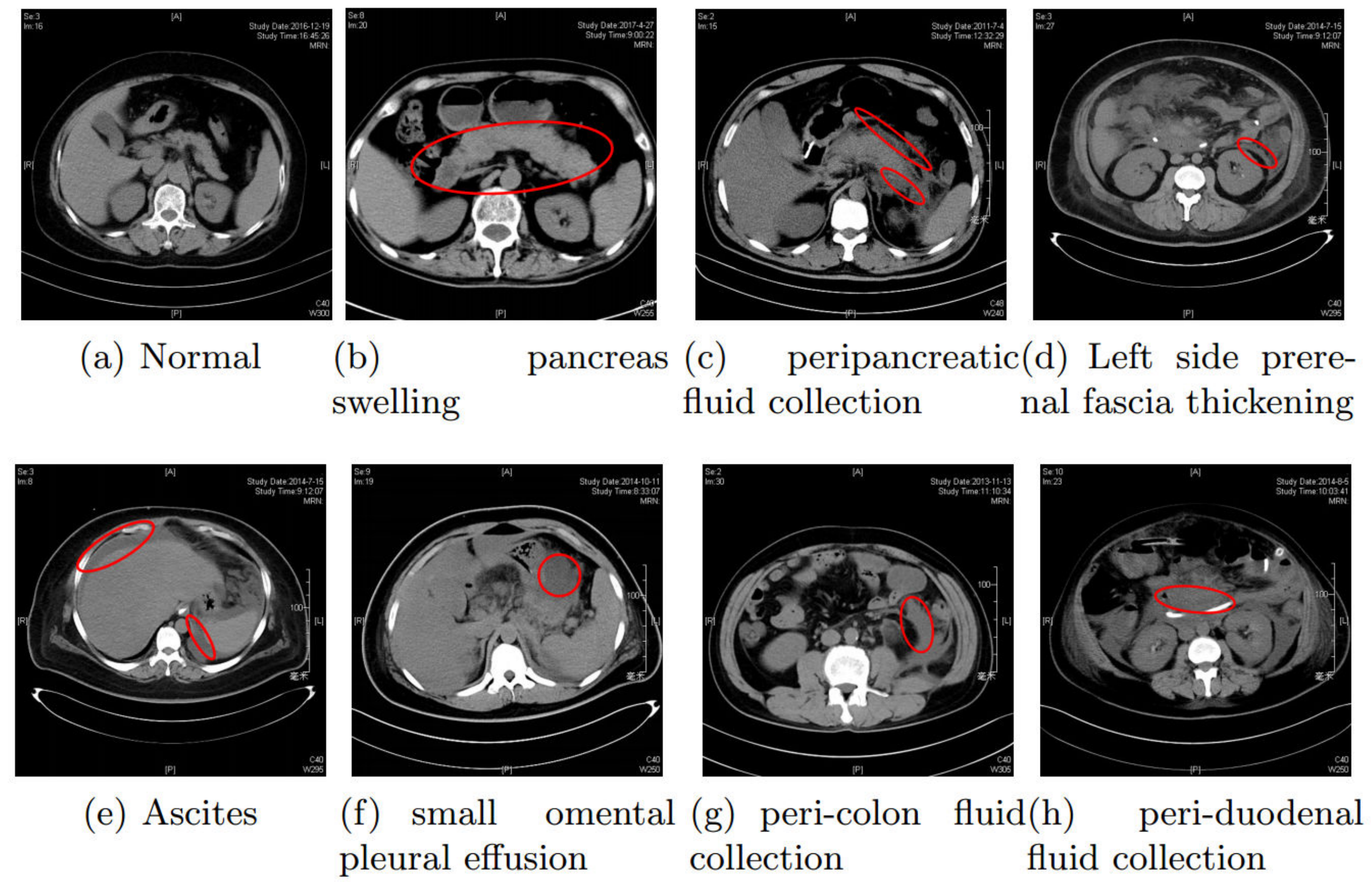}
\end{center}
%  \vspace{2mm}
 \caption{Comparisons of different abdominal CT scan images.(a) normal abdominal CT scan image. (b) pancreas swelling. (c) peripancreatic fluid collection. (d) left side prerenal fascia thickening. (e) ascites. (f)small omental pleural effusion. (g) peri-colon fluid collection. (h) peri-duodenal fluid collection.}
 \label{fig:all}
\end{figure}

  \subsection{Database collection}

  Since there is no public dataset used to evaluate the performance for the pancreatitis recognition task, we collect abdominal CT scan images with the size of $512 \times 512$ from hospital and build a database of Abdominal CT Images about Pancreatitis (ACIP). ACIP consists of 2500 normal abdominal CT images and 2500 abdominal CT images of pancreatitis patients. Three doctors with years of clinical experience are invited to annotate the images. Each image is annotated with 0 or 1, where 0 denotes a normal abdominal CT image and 1 denotes an abdominal CT image of pancreatitis patients. To learn and test models, this ACIP database is divided into two parts: 80\% for training and 20\% for testing.%which is shown in Table 2.

  \begin{table}[htbp]
  	\centering
  	\caption{ The number of images on the ACIP database. There are 4000 training images and 1000 testing images.}
  	%    \begin{tabular}{cc|l|rrrr}
  	\begin{tabular}{|c|c|c|}
  		\hline                     %& category  \multirow{1}[0]{*}{SVM}& Baseline \multicolumn{1}{c}{precision}
  		\multicolumn{1}{|c|}{Database} & normal  & pancreatitis \\
  		\hline
  		\hline
  		$\sharp$ training images  &  2000     &   2000         \\
  		\hline
  		$\sharp$	testing images   &  500      &   500 \\
    	\hline
  		$\sharp$	total   &  2500      &   2500 \\  		
  		\hline
  	\end{tabular}%
  	\label{tab:addlabel_1}%
  \end{table}%

  To better annotate CT images, we first analyze the abdominal CT scan images from pancreatitis patients and healthy people, and observe that the pancreatitis is determined by the small lesion area rather than most of areas as shown in Figure 4. For most of the abdominal CT images of pancreatitis patients, the lesion area mainly has the following characteristics: pancreas swelling, peripancreatic fluid collection, left side prerenal fascia thickening, ascites, small omental pleural effusion, peri-colon fluid collection, and peri-duodenal fluid collection. Our doctors give the results of diagnosis based on the above characteristics.

  \subsection{Experimental settings}
  To show the effectiveness of the proposed method for pancreatitis recognition, we compare it with the shallow method (i.e., SVM~\cite{suykens1999least}), and the classic deep learning models (i.e., AlexNet~\cite{krizhevsky2012imagenet}, VGG11~\cite{simonyan2014very} and VGG16~\cite{simonyan2014very}). For SVM, we reduce the size of the input image from $512\times512$ to $224\times224$ and extract features by SVD.

  In experiments, the following three baselines are also implemented to illustrate the effectiveness of the proposed region-manipulated scheme in RMFN:
(1)\textbf{Original CNN} (i.e., VGG11, VGG16, and AlexNet). The original CNN directly extracts features from the whole image and obtains the global information. This baseline can illustrate the importance of deep features. (2)\textbf{M-RMNF-A} (M is VGG11, VGG16, or AlexNet). This baseline extracts features from the whole image and small local areas at Scale 1. (3)\textbf{M-RMNF-B} (M is VGG11, VGG16, or AlexNet). This baseline extracts features from the whole image and big local areas at Scale 2.  Accordingly, for such three types of deep networks, the proposed RMFN also has three versions, termed as M-RMFN-C (i.e., VGG11-RMFN-C, VGG16-RMNF-C, and AlexNet-RMFN-C). The evaluation criterions include the recognition accuracy, precision, recall and F1 score.

\subsection{Results and analysis}

   Table 2 lists the results of 12 different networks on the collected ACIP database in terms of recognition accuracy, precision, recall and F1 on the ACIP database. We can see that AlexNet-RMFN-C achieves the best performance.
   For convenient comparison, we put the model with the same base networks into a family. For each family, the proposed M-RMFN-C achieves better performance than the other baselines. In particular, for the VGG16 family, VGG16-RMFN-C improves about 4.1\% compared with the original VGG16. It is indicated M-RMFN-C capturing the multi-scale local information is superior to M-RMFN-A and M-RMFN-B targeting only single-scale local information.
\begin{table*}[htbp]
%  \scriptsize
  \centering
  \caption{Performance comparison with different baseline networks on the ACIP database.}
%    \begin{tabular}{cc|l|rrrr}
    \begin{tabular}{|c|c|cccc|}
    \hline                     %& category  \multirow{1}[0]{*}{SVM}& Baseline
    \multicolumn{1}{|c|}{model}  & \multicolumn{1}{c|}{Scale} & \multicolumn{1}{c}{precision($\%$)} & \multicolumn{1}{c}{recall($\%$)} & \multicolumn{1}{c}{F1($\%$)} & \multicolumn{1}{c|}{accuracy($\%$)} \\
    \hline
    \hline
    %\multirow{13}[0]{*}{PACT}                   $ \%$
            SVM       &   $224\times224$    &  $79.6$ & $78.6$ & $79.1$ & $78.6$ \\             %\multicolumn{2}{|c|}{SVM}
    %\cline{1-6}
    \hline
            VGG11          &   $224\times224$                      & $87.8$  & $96.0$  & $91.7$  & $91.3$  \\  %\multirow{4}[0]{*}{Vgg11} &           &&&
            VGG11-RMFN-A   &   $224\times224+32\times32$           & $88.4$  & $96.0$  & $92.0$  & $91.7$  \\
            VGG11-RMFN-B   &   $224\times224+48\times48$           & $89.1$  & $94.8$  & $91.9$  & $91.6$   \\
            VGG11-RMFN-C   &   $224\times224+48\times48+32\times32$& $90.2$  & $97.4$  & $93.7$  & \bm{$93.4$}  \\
%    \cmidrule(r){2-7}
    \hline%\cline{1-6}
            VGG16        &   $224\times224$                        & $87.9$  & $94.2$  & $90.9$  & $90.6$\\  %\multirow{4}[0]{*}{Vgg16} &       &&&
            VGG16-RMFN-A &   $224\times224+32\times32$             & $91.4$  & $97.8$  & $94.5$  & $94.3$   \\
            VGG16-RMFN-B &   $224\times224+48\times48$             & $90.3$  & $97.2$  & $93.6$  & $93.4$   \\
            VGG16-RMFN-C &   $224\times224+48\times48+32\times32$  & $92.1$  & $97.8$  & $94.9$  & \bm{$94.7$}   \\
%    \cmidrule(r){2-7}
    \hline%\cline{1-6}
            AlexNet        & $224\times224$                        & $92.8$  & $98.6$  & $95.6$  & $95.5$   \\  %\multirow{4}[0]{*}{AlexNet} &       &&&
            AlexNet-RMFN-A & $224\times224+36\times36$             & $93.7$  & $98.4$  & $96.0$  & $95.9$   \\
            AlexNet-RMFN-B & $224\times224+48\times48$             & $94.1$  & $98.8$  & $96.4$  & $96.3$   \\
            AlexNet-RMFN-C & $224\times224+48\times48+36\times36$  & $95.3$  & $98.2$  & $96.7$  & \bm{$96.7$}   \\
    \hline
    \end{tabular}%
  \label{tab:addlabe2}%
\end{table*}%

  To further illustrate our networks architecture better, we visualize the learned features to intuitively compare the proposed RMFN and baselines, as shown in Figure 5. We visualize feature maps of final convolutional layers and added the feature map to original image by suitable weights. Figure 5(a) shows input CT scan images of pancreatitis, where the areas of pancreatic fluid are outlined in red circles. Figure 5(b), Figure 5(c), Figure 5(d) and Figure 5(e) are the visualized features learned from AlexNet, AlexNet-RMFN-C, VGG16, and VGG16-RMFN-C, respectively. Comparing with the features of AlexNet and VGG16, the proposed AlexNet-RMFN-c and VGG16-RMFN-C locate the lesion areas more accurately.

\section{Conclusion}
\label{}
 % {\color{blue}
  In this work, we proposed a novel Region-Manipulated Fusion Networks (RMFN) architecture to recognize pancreatitis on the abdominal CT scan images by capturing the key characteristics of the local lesion. In particular, a novel region-manipulated scheme in RMFN can effectively highlight the imperceptible lesion regions by aggregating the multi-scale local information into feature maps. This region-manipulated scheme can enhance the local discriminative region, meanwhile, weaken the non-discriminative regions. To evaluate the performance of the propose method,  we build a high-quality abdominal CT image database with/without pancreatitis. Finally, we conduct experiments to demonstrate the effectiveness of the proposed method for pancreatitis recognition. In future work, we will push the region-manipulated idea into the powerful Generative Adversarial Net and further improve the performance of pancreatitis recognition.%}

\bibliographystyle{IEEEtran}
\bibliography{paper}

\begin{thebibliography}{10}
\providecommand{\url}[1]{#1}
\csname url@samestyle\endcsname
\providecommand{\newblock}{\relax}
\providecommand{\bibinfo}[2]{#2}
\providecommand{\BIBentrySTDinterwordspacing}{\spaceskip=0pt\relax}
\providecommand{\BIBentryALTinterwordstretchfactor}{4}
\providecommand{\BIBentryALTinterwordspacing}{\spaceskip=\fontdimen2\font plus
\BIBentryALTinterwordstretchfactor\fontdimen3\font minus
  \fontdimen4\font\relax}
\providecommand{\BIBforeignlanguage}[2]{{%
\expandafter\ifx\csname l@#1\endcsname\relax
\typeout{** WARNING: IEEEtran.bst: No hyphenation pattern has been}%
\typeout{** loaded for the language `#1'. Using the pattern for}%
\typeout{** the default language instead.}%
\else
\language=\csname l@#1\endcsname
\fi
#2}}
\providecommand{\BIBdecl}{\relax}
\BIBdecl

\bibitem{yadav2013epidemiology}
D.~Yadav and A.~B. Lowenfels, ``The epidemiology of pancreatitis and pancreatic
  cancer,'' \emph{Gastroenterology}, vol. 144, no.~6, pp. 1252--1261, 2013.

\bibitem{banks2013classification}
P.~A. Banks, T.~L. Bollen, C.~Dervenis, H.~G. Gooszen, C.~D. Johnson, M.~G.
  Sarr, G.~G. Tsiotos, and S.~S. Vege, ``Classification of acute
  pancreatitis—2012: revision of the atlanta classification and definitions
  by international consensus,'' \emph{GUT}, vol.~62, no.~1, pp. 102--111, 2013.

\bibitem{cavestro2015single}
G.~M. Cavestro, G.~Leandro, M.~Di~L., R.~A. Zuppardo, O.~B. Morrow,
  C.~Notaristefano, G.~Rossi, S.~G.~G. Testoni, G.~Mazzoleni, M.~Alessandri
  \emph{et~al.}, ``A single-centre prospective, cohort study of the natural
  history of acute pancreatitis,'' \emph{Digestive and Liver Disease}, vol.~47,
  no.~3, pp. 205--210, 2015.

\bibitem{nawaz2013revised}
H.~Nawaz, R.~Mounzer, D.~Yadav, J.~G. Yabes, A.~Slivka, D.~C. Whitcomb, and
  G.~I. Papachristou, ``Revised atlanta and determinant-based classification:
  application in a prospective cohort of acute pancreatitis patients,''
  \emph{AJG}, vol. 108, no.~12, p. 1911, 2013.

\bibitem{girshick2015fast}
R.~Girshick, ``Fast r-cnn,'' in \emph{ICCV}, 2015.

\bibitem{krizhevsky2012imagenet}
A.~Krizhevsky, I.~Sutskever, and G.~E. Hinton, ``Imagenet classification with
  deep convolutional neural networks,'' in \emph{NIPS}, 2012.

\bibitem{sun2015deeply}
Y.~Sun, X.~Wang, and X.~Tang, ``Deeply learned face representations are sparse,
  selective, and robust,'' in \emph{CVPR}, 2015, pp. 2892--2900.

\bibitem{gulshan2016development}
V.~Gulshan, L.~Peng, M.~Coram, M.~C. Stumpe, D.~Wu, A.~Narayanaswamy,
  S.~Venugopalan, K.~Widner, T.~Madams, J.~Cuadros \emph{et~al.}, ``Development
  and validation of a deep learning algorithm for detection of diabetic
  retinopathy in retinal fundus photographs,'' \emph{JAMA}, vol. 316, no.~22,
  pp. 2402--2410, 2016.

\bibitem{esteva2017dermatologist}
A.~Esteva, B.~Kuprel, R.~A. Novoa, J.~Ko, S.~M. Swetter, H.~M. Blau, and
  S.~Thrun, ``Dermatologist-level classification of skin cancer with deep
  neural networks,'' \emph{Nature}, vol. 542, no. 7639, pp. 115--118, 2017.

\bibitem{golan2016lung}
R.~Golan, C.~Jacob, and J.~Denzinger, ``Lung nodule detection in ct images
  using deep convolutional neural networks,'' in \emph{IJCNN}, 2016.

\bibitem{lecun2015deep}
Y.~LeCun, Y.~Bengio, and G.~Hinton, ``Deep learning,'' \emph{Nature}, vol. 521,
  no. 7553, pp. 436--444, 2015.

\bibitem{zhou2016learning}
B.~Zhou, A.~Khosla, A.~Lapedriza, A.~Oliva, and A.~Torralba, ``Learning deep
  features for discriminative localization,'' in \emph{CVPR}.\hskip 1em plus
  0.5em minus 0.4em\relax IEEE, 2016, pp. 2921--2929.

\bibitem{parkhi2015deep}
O.~M. Parkhi, A.~Vedaldi, A.~Zisserman \emph{et~al.}, ``Deep face
  recognition.'' in \emph{BMVC}, 2015.

\bibitem{liu2017sphereface}
W.~Liu, Y.~Wen, Z.~Yu, M.~Li, B.~Raj, and L.~Song, ``Sphereface: Deep
  hypersphere embedding for face recognition,'' in \emph{CVPR}, 2017, pp.
  212--220.

\bibitem{akata2014good}
Z.~Akata, F.~Perronnin, Z.~Harchaoui, and C.~Schmid, ``Good practice in
  large-scale learning for image classification,'' \emph{IEEE TPAMI}, vol.~36,
  no.~3, pp. 507--520, 2014.

\bibitem{wang2017residual}
F.~Wang, M.~Jiang, C.~Qian, S.~Yang, C.~Li, H.~Zhang, X.~Wang, and X.~Tang,
  ``Residual attention network for image classification,'' in \emph{CVPR},
  2017, pp. 3156--3164.

\bibitem{szegedy2015going}
C.~Szegedy, W.~Liu, Y.~Jia, P.~Sermanet, S.~Reed, D.~Anguelov, D.~Erhan,
  V.~Vanhoucke, and A.~Rabinovich, ``Going deeper with convolutions,'' in
  \emph{CVPR}, 2015.

\bibitem{ren2015faster}
S.~Ren, K.~He, R.~Girshick, and J.~Sun, ``Faster r-cnn: Towards real-time
  object detection with region proposal networks,'' in \emph{NIPS}, 2015.

\bibitem{chen2017deeplab}
L.-C. Chen, G.~Papandreou, I.~Kokkinos, K.~Murphy, and A.~L. Yuille, ``Deeplab:
  Semantic image segmentation with deep convolutional nets, atrous convolution,
  and fully connected crfs,'' \emph{IEEE TPAMI}, vol.~40, no.~4, pp. 834--848,
  2017.

\bibitem{badrinarayanan2017segnet}
V.~Badrinarayanan, A.~Kendall, and R.~Cipolla, ``Segnet: A deep convolutional
  encoder-decoder architecture for image segmentation,'' \emph{IEEE TPAMI},
  vol.~39, no.~12, pp. 2481--2495, 2017.

\bibitem{chen2018encoder}
L.-C. Chen, Y.~Zhu, G.~Papandreou, F.~Schroff, and H.~Adam, ``Encoder-decoder
  with atrous separable convolution for semantic image segmentation,'' in
  \emph{ECCV}, 2018, pp. 801--818.

\bibitem{simonyan2014very}
K.~Simonyan and A.~Zisserman, ``Very deep convolutional networks for
  large-scale image recognition,'' \emph{arXiv}, 2014.

\bibitem{he2016deep}
K.~He, X.~Zhang, S.~Ren, and J.~Sun, ``Deep residual learning for image
  recognition,'' in \emph{CVPR}, 2016.

\bibitem{ren2018adversarial}
J.~Ren, I.~Hacihaliloglu, E.~A. Singer, D.~J. Foran, and X.~Qi, ``Adversarial
  domain adaptation for classification of prostate histopathology whole-slide
  images,'' in \emph{MICCAI}.\hskip 1em plus 0.5em minus 0.4em\relax Springer,
  2018, pp. 201--209.

\bibitem{zhang2018skin}
J.~Zhang, Y.~Xie, Q.~Wu, and Y.~Xia, ``Skin lesion classification in dermoscopy
  images using synergic deep learning,'' in \emph{MICCAI}.\hskip 1em plus 0.5em
  minus 0.4em\relax Springer, 2018, pp. 12--20.

\bibitem{liao2019evaluate}
F.~Liao, M.~Liang, Z.~Li, X.~Hu, and S.~Song, ``Evaluate the malignancy of
  pulmonary nodules using the 3-d deep leaky noisy-or network,'' \emph{IEEE
  TNNLS}, 2019.

\bibitem{zhu2018deeplung}
W.~Zhu, C.~Liu, W.~Fan, and X.~Xie, ``Deeplung: Deep 3d dual path nets for
  automated pulmonary nodule detection and classification,'' in
  \emph{WACV}.\hskip 1em plus 0.5em minus 0.4em\relax IEEE, 2018, pp. 673--681.

\bibitem{farag2017bottom}
A.~Farag, L.~Lu, H.~R. Roth, J.~Liu, E.~Turkbey, and R.~M. Summers, ``A
  bottom-up approach for pancreas segmentation using cascaded superpixels and
  (deep) image patch labeling,'' \emph{IEEE TIP}, vol.~26, no.~1, pp. 386--399,
  2017.

\bibitem{man2019deep}
Y.~Man, Y.~Huang, J.~F.~X. Li, and F.~Wu, ``Deep q learning driven ct pancreas
  segmentation with geometry-aware u-net,'' \emph{IEEE TMI}, 2019.

\bibitem{yu2018recurrent}
Q.~Yu, L.~Xie, Y.~Wang, Y.~Zhou, E.~K. Fishman, and A.~L. Yuille, ``Recurrent
  saliency transformation network: Incorporating multi-stage visual cues for
  small organ segmentation,'' in \emph{CVPR}, 2018, pp. 8280--8289.

\bibitem{lambin2012radiomics}
P.~Lambin, E.~Rios-Velazquez, R.~Leijenaar, S.~Carvalho, R.~G. van Stiphout,
  P.~Granton, C.~M. Zegers, R.~Gillies, R.~Boellard, A.~Dekker \emph{et~al.},
  ``Radiomics: extracting more information from medical images using advanced
  feature analysis,'' \emph{EJC}, vol.~48, no.~4, pp. 441--446, 2012.

\bibitem{gao2016holistic}
M.~Gao, U.~Bagci, L.~Lu, A.~Wu, M.~Buty, H.~C. Shin, H.~Roth, G.~Z. Papadakis,
  A.~Depeursinge, R.~M. Summers \emph{et~al.}, ``Holistic classification of ct
  attenuation patterns for interstitial lung diseases via deep convolutional
  neural networks,'' \emph{CMBBE: Imaging \& Visualization}, pp. 1--6, 2016.

\bibitem{armato2011lung}
S.~G. Armato, G.~McLennan, L.~Bidaut, M.~F. McNitt-Gray, C.~R. Meyer, A.~P.
  Reeves, B.~Zhao, D.~R. Aberle, C.~I. Henschke, E.~A. Hoffman \emph{et~al.},
  ``The lung image database consortium (lidc) and image database resource
  initiative (idri): a completed reference database of lung nodules on ct
  scans,'' \emph{Medical Physics}, vol.~38, no.~2, pp. 915--931, 2011.

\bibitem{setio2016pulmonary}
A.~A.~A. Setio, F.~Ciompi, G.~Litjens, P.~Gerke, C.~Jacobs, S.~J. van R.,
  M.~M.~W. Wille, M.~Naqibullah, C.~I. S{\'a}nchez, and B.~van G., ``Pulmonary
  nodule detection in ct images: false positive reduction using multi-view
  convolutional networks,'' \emph{IEEE TMI}, vol.~35, no.~5, pp. 1160--1169,
  2016.

\bibitem{shin2016deep}
H.~C. Shin, H.~R. Roth, M.~Gao, L.~Lu, Z.~Xu, I.~Nogues, J.~Yao, D.~Mollura,
  and R.~M. Summers, ``Deep convolutional neural networks for computer-aided
  detection: Cnn architectures, dataset characteristics and transfer
  learning,'' \emph{IEEE TMI}, vol.~35, no.~5, pp. 1285--1298, 2016.

\bibitem{gao2015automatic}
X.~Gao, S.~Lin, and T.~Y. Wong, ``Automatic feature learning to grade nuclear
  cataracts based on deep learning,'' \emph{IEEE TBE}, vol.~62, no.~11, pp.
  2693--2701, 2015.

\bibitem{kamnitsas2017efficient}
K.~Kamnitsas, C.~Ledig, V.~F. Newcombe, J.~P. Simpson, A.~D. Kane, D.~K. Menon,
  D.~Rueckert, and B.~Glocker, ``Efficient multi-scale 3d cnn with fully
  connected crf for accurate brain lesion segmentation,'' \emph{Medical image
  analysis}, vol.~36, pp. 61--78, 2017.

\bibitem{suykens1999least}
J.~A. Suykens and J.~Vandewalle, ``Least squares support vector machine
  classifiers,'' \emph{NPL}, vol.~9, no.~3, pp. 293--300, 1999.

\end{thebibliography}

\end{document}